\documentclass[traditabstract]{aa}

\usepackage{natbib}
\bibpunct{(}{)}{;}{a}{}{,} % to follow the A&A style
\usepackage{color}
\usepackage{graphicx}
\usepackage{amsmath}

\begin{document}
\renewcommand{\vec}[1]{\bmath{#1}}
\newcommand{\sign}{\textrm{sign}}
\newcommand{\pd}[2]{\frac{\partial #1}{\partial #2}}
\newcommand{\DS}{\displaystyle}
\newcommand{\HALF}{\frac{1}{2}}
\newcommand{\mathi}{\rm i}

\renewcommand{\vec}[1]{\mathbf{#1}}
\newcommand{\hvec}[1]{\hat{\mathbf{#1}}}
\newcommand{\av}[1]{\left<#1\right>}
\newcommand{\red}[1]{\color{red} #1 \color{black}}
\newcommand{\blue}[1]{\color{blue} #1 \color{black}}
\newcommand{\ac}[1]{\textcolor{red}{#1}}

\title{Making Fanaroff-Riley I radio sources} \subtitle{III. The effects of the magnetic field on relativistic jets' propagation and source morphologies}

\author{S. Massaglia\inst{1}, G. Bodo\inst{2}, P. Rossi\inst{2},
  A. Capetti\inst{2}, A. Mignone\inst{1}}

\authorrunning{S. Massaglia et al.}
\titlerunning{Making Fanaroff-Riley I radio sources.}

\institute{ Dipartimento di Fisica, Universit\`a degli Studi di Torino, via
  Pietro Giuria 1, 10125 Torino, Italy \and INAF/Osservatorio Astrofisico di
  Torino, via Osservatorio 20, 10025 Pino Torinese, Italy}

\date{Received ?? / Accepted ??}

%\pagerange{\pageref{firstpage}--\pageref{lastpage}} \pubyear{2007}

\titlerunning{}

\label{firstpage}

\abstract{Extragalactic radio sources appear under different morphologies, the most frequent ones are classified as Fanaroff-Riley type I (FR I), typically with lower luminosities, and Fanaroff-Riley type II, (FR II), typically more luminous. This simple classification, however, has many exceptions that we intend to investigate. Following previous analyses in the three-dimensional Hydrodynamic and Magneto-Hydrodynamic limits, we extend the numerical investigation to the  Relativistic Magneto-Hydrodynamic regime, to include sources whose jet kinetic power sets in the range that separates FR Is from FR IIs.

We consider weakly and mildly relativistic, underdense, supersonic jets that propagate in a stratified medium. In the model, the ambient temperature increases with distance from the jet origin maintaining constant pressure. 
We present three cases with low, high and intermediate kinetic luminosity that evolve  into different morphologies. We find that the resulting morphology can be highly time dependent and that, apart from the jet power, the jet-to-ambient density ratio and the magnetization parameter play a crucial role in the jet evolution as well.}

\maketitle

%%%%%%%%%%%%%%%%%%%%%%%%%%%%%%%%%%%%%%%%%%%%%%%%%%%
\section{Introduction}
%
%
%
%%%%%%%%%%%%%%%%%%%%%%%%%%%%%%%%%%%%%%%%%%%%%%%%%%%
Extragalactic radio sources present variegated morphologies and in addition to the well-known Fanaroff-Riley type I (FR~I) and II  (FR~II) sources, other more distorted morphological subclasses have been defined: sources with narrow angle and wide angle tails \citep{owen76, rudnick77} in which diffuse plumes are either smoothly or sharply bent with respect to the initial jet direction, or the so called S-, X- and Z-shaped sources, where the distortion affects the radio lobes (see, e.g., \citealt{leahy84}). The origin of these morphologies can be related to the various physical parameters describing their jets and/or the external medium in which they expand. To improve the general understanding of the Active Galactic Nuclei (AGN) jet properties and to attempt to reproduce these different morphologies by numerical simulations (see a recent review by \cite{Kompor21}), 
several authors have recently examined the problem in the limit of the hydrodynamic (HD)  up to the relativistic magneto-hydrodynamic (RMHD) extension \citep{smith19,wein17,ehlert18,yuton18,davel19,vdwest19,peru20,Mig2010,mukh20}.  With this aim, \cite{massaglia16} (Paper I) examined the evolution of supersonic HD jets of different power by means of 3D simulations. They found that FR~I or FR~II morphology can be obtained depending on the jet parameters and, in particular, on the jet power; moreover, 2D simulations cannot capture the transition between the two morphologies.  \citet{massaglia19}  (Paper II) extended the analysis to MHD simulations of low power jets, showing that,  an increase in the strength of magnetic field  leads to a strong wiggling of the jet, which eventually causes its fragmentation with the formation of one or more strong shocks. After these shocks the flow acquires a turbulent more diffuse structure and it moves at a much lower velocity. The jet structure observed in these simulations is indeed very reminiscent of the wide angle tail (WAT) class of radio sources.

In the present paper we extend the analysis to the   RMHD regime by increasing the simulated jet velocity with respect to Papers I and II. We examine jets with kinetic powers that differ by about an order of magnitude   across the values that characterize the transition between FR~I and FR~II sources, varying the jet-to-ambient density ratio, the plasma-$\beta$ parameter and the Lorentz factor and we follow the jet temporal evolution towards a quasi steady state. Our goal is thus to answer the questions related to i) the role of the various parameters in determining the jet evolution and its morphology, and ii) about the importance of the observation epoch for classifying the object. We do this by carrying out numerical simulations in three spatial dimensions and exploring the parameter space mentioned above.

The paper is organized as follows: in Section 2 we write the RMHD equations, set the initial and boundary conditions and constrain the physical parameters of the problem. In Section 3, we present the results that are discussed in Section 4.

%%%%%%%%%%%%%%%%%%%%%%%%%%%%%%%%%%%%%%%%%%%%%%%%%%%%%%%%%%%%%%%%%%%%%%%%
\section{Problem description}
%
%
%
%%%%%%%%%%%%%%%%%%%%%%%%%%%%%%%%%%%%%%%%%%%%%%%%%%%%%%%%%%%%%%%%%%%%%%%%

\subsection{Relativistic MHD equations and numerical approach}
Numerical simulations are carried out by solving the equations of relativistic ideal magneto-hydrodynamics in 3D.
They are expressed by
\begin{equation}
 \partial_t \left( \gamma \rho \right) + \nabla \cdot \left(\gamma \rho \vec{v}\right)= 0,
  \label{eq:RMHD_rho} \\ 
  %\noalign{\smallskip}
\end{equation}
\begin{equation}
%\begin{array}{ll}
\partial_t \left( \gamma^2 w \vec {v} + \vec {E} \times \vec {B} \right) + 
\nonumber
\end{equation}
\begin{equation}
\nabla \cdot 
\left( \gamma^2 w \ \vec {v} \vec {v} - \vec {E} \vec {E} - \vec {B} \vec {B} + (p+u_{\mathrm {em}})\vec{I} \right)= 0,
\label{eq:RMHD_mom} \\ 
  %\noalign{\smallskip}
%\end{array}
\end{equation}
\begin{equation}
 \partial_t \left( \gamma^2 w -p + u_{\mathrm {em}} \right) + \nabla \cdot \left( \gamma^2 w \vec {v} + \vec {E} \times \vec {B} \right)= 0,
\label{eq:RMHD_eng}
\end{equation}
\begin{equation} \label{eq:ind}
\partial_t \vec {B} + \nabla \times \vec {E}=0
\end{equation}
and the electric field $\vec{E}$ is provided by the ideal condition $\vec{E}+\vec{v}\times \vec{B}=0$.
In this set of equations, $\rho$ is the rest number density, $\vec{v}$ is the  velocity three-vector (in units of the light speed $c$),  $\vec{B}$ is the laboratory magnetic field (in units of $\sqrt{4 \pi}$), $\gamma=1/\sqrt{1-v^2}$, $w=e+p$, $u_{\rm {em}}=(E^2+B^2)/2$, and $\vec{I} $ is the unit $3 \times 3$ tensor. 
We assume a constant $\Gamma$-law equation of state with $\Gamma = 5/3$ being the specific heat ratio.

Eqns. (\ref{eq:RMHD_rho})-(\ref{eq:ind}) were solved using the ideal RMHD module of the PLUTO code \citep{PLUTO}, based on shock-capturing Godunov-type methods.
Equations were evolved in conservative form with a second-order Runge-Kutta time stepping, a linear reconstruction and the HLLD Riemann solver \citep{MigUglBod.2009}, constrained transport \citep{Balsara99, Londrillo04} was used to maintain the condition $\nabla \cdot \vec{B} = 0$.
As in Papers I and II, 3D simulations were carried out on a Cartesian domain with coordinates in the range $x\in [-L_x/2,L_x/2]$, $y\in [0,L_y]$ and $z\in [-L_z/2,L_z/2]$ (lengths are expressed in units of the jet radius). 
Here $L_x=L_z$, while we chose $y$ as the jet propagation direction.
We employed a grid of $N_x\times N_y\times N_z$ zones with a uniform grid spacing along the $y$  (axial) direction.
In the transverse direction, the region $x,y\in[-6,6]$ was covered with 100 uniform zones, while a geometrically stretched grid was used outside  this area.
The actual domain size and grid resolution are reported in Table \ref{labvalues} for the different cases discussed in this work.

%%%%%%%%%%%%%%%%%%%%%%%%%%%%%%%%%%%%%%%%%%%%%%%%%%%%%%%%%%%%%%%%%%%%%%%%
\subsection{Initial and boundary conditions and physical parameters}
\label{sec:theory}
%%%%%%%%%%%%%%%%%%%%%%%%%%%%%%%%%%%%%%%%%%%%%%%%%%%%%%%%%%%%%%%%%%%%%%%%
  
\begin{table*}[tb] 
\begin{center} 
\caption{Parameter set used in the numerical simulations.} 
\begin{tabular}{lccclcccccll} 
\hline\\[-12pt]
 Case    & $\eta$ & $\gamma_{\mathrm j}$ & $V_{\mathrm j}$ &  $M$
         & ${\cal L}_{\mathrm {kin}} \ ({\mathrm {erg \ s^{-1}}})$
         & $\alpha$ & $\beta$
         & $L_x \times L_y \times L_z$
         & $N_x \times N_y \times N_z$ \\[3pt]

\hline\\[-12pt] 
A1 & $10^{-3}$         & 1.005 & 0.1 & 6  & $7\times 10^{42}$ & 1 & 0.5 & $80\times 200\times 80$   & $228\times 2000\times 228$  \\ 
A2 & $10^{-3}$         & 1.005 & 0.1 & 6  & $7\times 10^{42}$ & 2 & 0.5&  $80\times 200\times 80$   & $228\times 1200\times 228$  \\ 
B & $10^{-3}$         & 1.05  & 0.3 & 19 & $2\times 10^{44}$ & 1 & 0.5& $100\times 600\times 100$ & $244\times 4800\times 244$  \\
C1 & $3\times 10^{-4}$ & 1.05  & 0.3 & 10 & $6\times 10^{43}$ & 1 &0.5   & $100\times 500\times 100$ & $244\times 4000\times 244$ \\
C2 & $3\times 10^{-4}$ & 1.05  & 0.3 & 10 & $6\times 10^{43}$ & 2  &$ 0.5$  & $100\times 500\times 100$ & $244\times 4000\times 244$ \\
C-RHD & $3\times 10^{-4}$ & 1.05  & 0.3 & 10 & $6\times 10^{43}$ & 1  &$ 10^6$  & $100\times 500\times 100$ & $244\times 4000\times 244$ \\
[3pt]
\hline 
\end{tabular} 
\label{labvalues} 
\end{center}
Columns 2-7 correspond to the physical parameters $\eta$,  $\gamma$, $V_j$, $M$, ${\cal L}_{\mathrm{kin}}$, $\alpha$ and $\beta$ described in the text, respectively.
%, while ${\cal L}_{\mathrm R}$ ($7^{\rm th}$ column) gives radio power estimated from \cite{cavagn10} (Appendix A.1).
The last two columns give the domain size and grid resolution for each case. 
\end{table*}

At $t=0$, the domain was filled with a perfect gas at rest,  unmagnetized, and with uniform pressure, but it was spherically stratified following a King-like profile:
\begin{equation}\label{eq:king}
  \rho(R)=\frac{\rho_{\mathrm c}}{1+\left(R/r_{\mathrm c}\right)^\alpha}
\end{equation}
 where $R$ is the spherical radius and $r_{\mathrm c}$ is the galactic core radius. We  set $\alpha=1$ for most of the cases examined except for the cases A2 and C2, where we will examine the case $\alpha=2$ as well for a comparison.  We note that the assumption of uniform pressure and a decreasing density leads to a temperature that increases with radius. In fact, X-ray observations show that in the galactic core the ambient temperature attains values $\sim 0.3-1$ keV and particle densities of a few particles per cm$^{-3}$, typical of the denser and cooler gas of the galactic core ($R<r_{\mathrm c}$) \citep{balm08}. When entering the intracluster medium, $R>r_{\mathrm c}$, the temperatures reach values of about $2-10$ keV and densities decline down to about $10^{-2}$ particles per cm$^{-3}$ \citep{vik06}. The behaviour of the overall temperature profile is then quite complex, the predominance of denser and cooler gas in the galactic core is likely due to radiative cooling, which is stronger in these conditions. The density profile in Eq. \ref{eq:king} and the corresponding temperature profile aims to capture the main properties of this behaviour when we assume $T_{\mathrm c} \sim 0.2$ keV and $\rho_{\mathrm c} = 1$ cm$^{-3}$ at $R=0$.

A cylindrical jet with velocity $V_j$, density $\rho_j$ and a purely azimuthal magnetic field was constantly injected from the boundary at $y=0$.  The jet at the injection boundary is in pressure equilibrium with the ambient, its configuration is axisymmetric and the equilibrium condition, in cylindrical coordinates, can be written as follows:
\begin{equation}\label{eq:equil}
\frac{dp}{dr}=-\frac{1}{2r^2} \frac{d(r^2 B_\phi^2)}{dr}
\end{equation}
where $r$ is the cylindrical radius and $B_{\phi}$ is the azimuthal component of the magnetic field. We  assumed the magnetic field corresponding to a constant current inside $r = a$ and zero outside:
\begin{equation}
B_\phi = \begin{cases}
  -B_{\mathrm m} r/a, &\text{if $r < a$;}\\
  -B_{\mathrm m} a/r, &\text{otherwise.}
\end{cases}
\end{equation}
In this way, in Eq. \ref{eq:equil} the term on the left-hand side can be integrated giving
\begin{equation}
p_{\mathrm j} =p_{\mathrm c}-B^2_{\mathrm m} {\mathrm {min}} \left( \frac{r^2}{a^2}, 1\right)
\end{equation}
where $p_{\mathrm j}$ is the jet gas pressure and $p_{\mathrm c}$ is a constant. Since we set the ambient pressure $p_{\mathrm a}$, it was convenient to specify $p_{\mathrm c}=p_{\mathrm a}+B_{\mathrm m}^2$, thus
\begin{equation}
p_{\mathrm j} =p_{\mathrm a}+B_{\mathrm m}^2 \left[1-{\mathrm {min}} \left( \frac{r^2}{a^2}, 1\right)\right] \,,
\end{equation}
and $p_{\mathrm j} =p_{\mathrm a}$ at $r=a$.

The magnetic field strength $B_{\mathrm m}$ was obtained from the definition of the plasma $\beta$-parameter,
\begin{equation}
  \beta=\frac{2 p_{\mathrm a}}{B_{\mathrm m}^2} \,.
\end{equation}

Outside the jet injection nozzle we imposed reflective boundary conditions. As we did in Paper II, to avoid  sharp  transitions,  we  smoothly  joined  the  injection  and reflective boundary values in the following way:
\begin{equation}\label{eq:bdprof}
    Q(x,z,t) = Q_r(x,z,t) + \frac{Q_j - Q_r(x,z,t)}{\cosh[(r/r_s)^n]}
\end{equation}
where $Q$ is  any  of the primitive flow variables, with the exception of jet velocity, which was replaced by the jet $y-$momentum. We note that $Q_r(x,z,t)$ are the corresponding time-dependent reflected values, while $Q_j$ are the constant injection values. In Eq. \ref{eq:bdprof} we set $r_s = 1$ and $n=6$ for all variables except  the density, for which we chose $r_s = 1.4$ and $n=8$. This choice ensures monotonicity in the momentum \citep{Massaglia96}.  On all the other computational boundaries, we imposed a condition of zero-gradient. 

In Papers I and II, and the present one, we examine the jet propagation  at the so-called VLA scale, when the jet's Lorentz factor has dropped from $\gamma \approx 5-10$, which is typical of the `VLBI scale' for both FR I and FR II jets \citep{gg01}, down to values close to the unity. The jet magnetic stability and the recollimation process at VLBI distances from the AGN has been recently examined by \cite{Matsum21}, and the jet deceleration of HD relativistic jets has been explored by \cite{rossi20}.

We non-dimensionalized the equations by using the velocity of light $c$ as the unit of velocity, the jet radius $r_j$ as the unit of length and $\rho_{\mathrm c}$ as the unit of density. 
To specify the problem completely,  besides  the  plasma-$\beta$ introduced above,  we need the following additional parameters: the Lorentz factor
\begin{equation}\label{eq:gamma}
  \gamma_{\mathrm j} = \frac{1}{\sqrt{1-V_{\mathrm j}^2}} \,, 
%\nonumber
\end{equation}
the density ratio
\begin{equation}\label{eq:eta}
  \eta = \frac{\rho_{\mathrm j}}{\rho_{\mathrm c}} \,,
\end{equation}
and the non-dimensional ambient pressure
\begin{equation}\label{eq:pi}
  \Pi= \frac{p_{\mathrm a}}{\rho_{\mathrm c} c^2} \,.
\end{equation}
For $\Pi$ we took the value $2 \times 10^{-7}$ constant for all the simulations, consistently with the assumed values of $\rho_{\mathrm c}$ and $T_{\mathrm c}$. Additionally, we have assumed the ratio between the core radius and the jet radius to be $r_{\mathrm c}/r_{\mathrm j} = 40$ for all the simulations.

In order to convert the non dimensional values to physical units, we had to make assumptions about the values for  units of length and density. As in our previous papers, for the (cylindrical) jet radius at the injection we assumed $r_{\mathrm j}=100$ pc, consequently, the galactic core radius results to be $r_{\mathrm c}=4$ kpc and the computational time unit  $\tau$, which is the light travel time over the initial jet radius, in physical units results $\tau \simeq 343$ yrs. 

In the limit of weakly relativistic and matter-dominated jets:
\begin{equation}
p_{\mathrm j}  / ( \rho_{\mathrm j} c^2) \ll 1 \,, \\
\end{equation}
and the kinetic jet power turns out to be:
\begin{equation}\label{eq:power}
  \begin{array}{rl}
{\cal L}_{\mathrm {kin}} \simeq & \pi  r_{\mathrm j}^2 \ \rho_{\mathrm j}  \ c^3 \  \gamma_{\mathrm j} ( \gamma_{\mathrm j}-1) \ V_{\mathrm j}
    \\
   \simeq & \DS {\cal L}_0 \times \left(\frac{r_{\mathrm j}}{100 {\mathrm pc}} \right)^2 
\eta \left( \frac{\rho_{\mathrm c}}{1 \ \mathrm {cm}^{-3}} \right)  \gamma_{\mathrm j} (\gamma_{\mathrm j}-1) V_{\mathrm j}  \, .
\end{array}
\end{equation}
where ${\cal L}_0 = 1.3 \times 10^{49} \quad {\mathrm {erg \ s^{-1}}}$.

An additional parameter of interest is the jet Mach number, which we define as
\begin{equation}\label{eq:Mach}
  M = \frac{V_{\mathrm j}}{c_{\mathrm {sj}}} \,,
\end{equation}
where 
%the jet velocity $V_{\mathrm j}$ and the sound speed $c_{\mathrm sj}$ are in unit of the speed of light $c$, and 
$c_{\mathrm sj}$ was computed at $r=a$. The value of $M$ is connected to $T_{\mathrm c}$. In fact, from the definition in Eq. \ref{eq:Mach} and the pressure equilibrium condition at $r=a$ we have the following:
\begin{equation}
M \simeq 3 \times 10^{6} \ V_{\mathrm j} \ \left(\frac{\eta} {T_{\mathrm c}}\right)^{1/2}.
\end{equation}
The maximum magnetic field intensity at injection is evaluated as follows:
\begin{eqnarray}
B_{\mathrm m} = B_0 \left( \frac{\rho_{\mathrm c}}{1 \ \mathrm {cm}^{-3}} \right)^{1/2}  \left( \frac{T_{\mathrm c}} {0.2 \mathrm{keV}} \right)^{1/2} \times \nonumber \\
 \left(\frac{\beta}{0.5} \right)^{-1/2} \ G \,,
 \label{eq:field}
\end{eqnarray}
where $B_0 \simeq 1.3 \times 10^{-4}$ G.

All the parameters of the simulations are reported in Table \ref{labvalues}.
In the presentation of the results, all quantities are expressed in physical units; their dimensional values, however, depend  on the choice of our unit of length $r_{\mathrm c} $ and density $\rho_{\mathrm c}$.
A different choice for these units would give different values for all the relevant physical quantities, since  lengths scale as $r_{\mathrm c}$, times also scale  as $r_{\mathrm c}$, pressure scales as $\rho_{\mathrm c}$ and the power and magnetic field strength scale as given in Eqs. \ref{eq:power}  and \ref{eq:field} respectively.

%%%%%%%%%%%%%%%%%%%%%%%%%%%%%%%%%%%%%%%%%%%%%%%%%%%
\section{Results}
%
%
%%%%%%%%%%%%%%%%%%%%%%%%%%%%%%%%%%%%%%%%%%%%%%%%%%%

As shown in Table \ref{labvalues}, we carried out six cases, with different values for the jet density, jet speed, and, consequently, different powers.
In particular, the jet parameters for cases A1 and A2 are the same and were chosen in connection with Paper II, with a jet velocity of $0.1c$. The difference between A1 and A2 is the density profile of the external medium, in case A2 we used the same profile as in Paper II ($\alpha =2$), while in case A1 we used a flatter profile ($\alpha = 1$).  For case B we increased the  power by about a factor 30, keeping the same density ratio, so the velocity became $0.3 c$. In cases C1 and C2, we kept the same velocity as case B, $V_j = 0.3c$, but we  lowered the  power by decreasing the density ratio by a factor $\sim 3$. As in  cases A1 and A2, the difference between  cases C1 and C2 is the density profile of the external medium, with $\alpha = 1$ in case C1 and $\alpha = 2$ in case C2. Finally case C-RHD has the same parameters as case C1, except the magnetization, which was considered to be very low, more precisely we considered  $\beta = 10^6$ and, essentially, we were dealing with an unmagnetized jet.  In all the other cases  the magnetization parameter $\beta$ has a constant value of $\beta = 0.5$\footnote{This value corresponds to $\beta = 3$ used in Paper II; in fact, therein, $\beta$ was computed from the average strength of magnetic field, while, for the sake of simplicity, here it is computed from the peak value.}, corresponding, for our reference units,  to a peak value of $B\approx 10^{-4}$ G.

\begin{figure*}[htp] 
\centering%
\includegraphics[width=1.8\columnwidth]{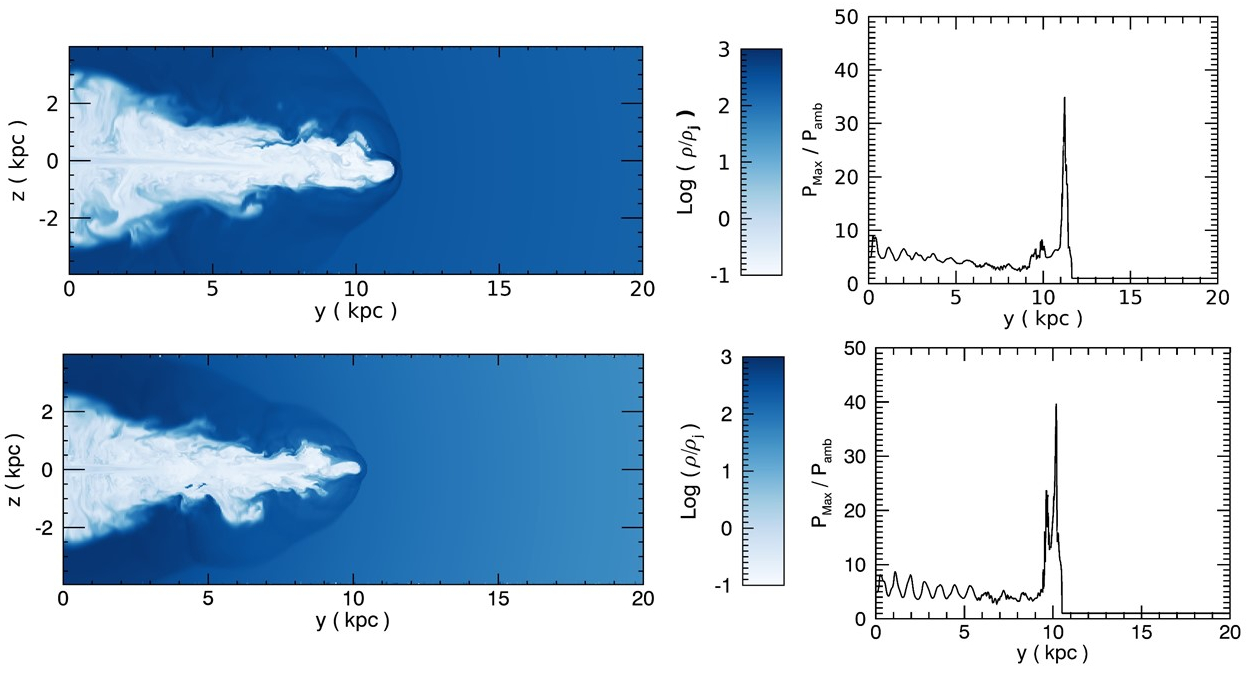}%

\caption{Top: Cut of the logarithmic density distribution for case A1  in the $(y,z)$ plane (left panel) and maximum gas pressure as a function of $y$ (right panel) at the time $t=1.9 \times 10^7$ yrs; Bottom:
Same as for case A2, but at the time  $t=1.3 \times 10^7$ yrs.
}
\label{fig:caseA1_1-2} 
\end{figure*}

\begin{figure*}[htp] 
\centering%
\includegraphics[width=1.6\columnwidth]{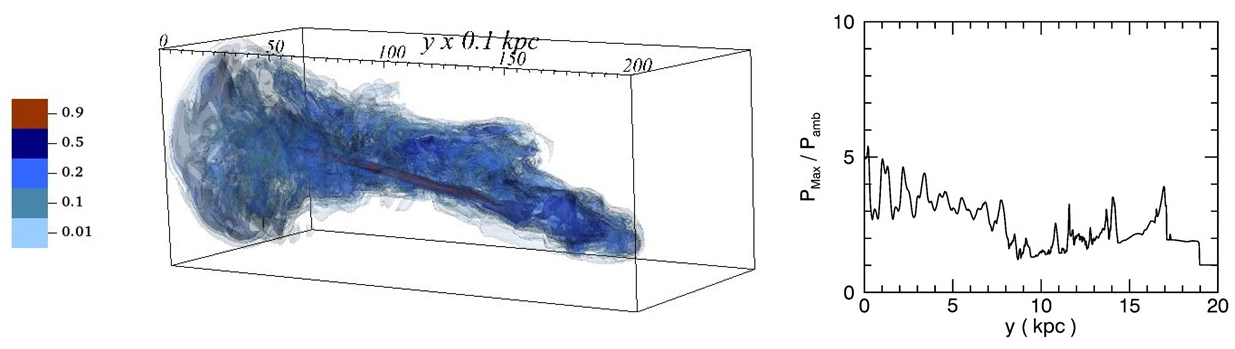}%

\caption{Left panel: 3D isocontours of the tracer distribution at a time $t=3.2\times 10^7$ yrs for case A1; the size of the computational box is $8 \times 20 \times 8$ kpc. Right panel: maximum gas pressure as a function of $y$.
}
\label{fig:caseA_1_tr} 
\end{figure*}

\begin{figure*}[htp] 
\centering%
\includegraphics[width=1.6\columnwidth]{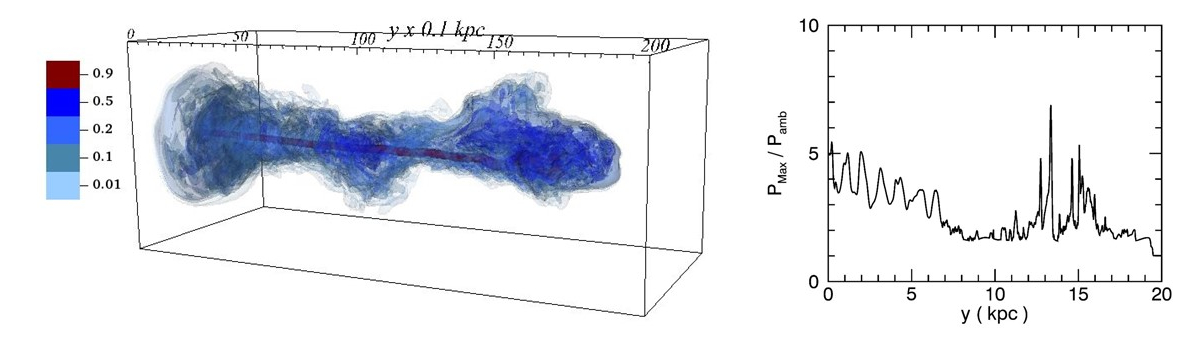}%
\caption{Left panel: 3D isocontours of the tracer distribution at a time $t=2.1\times 10^7$ yrs for case A2; the size of the computational box is $8 \times 20 \times 8$ kpc. Right panel: the maximum gas pressure as a function of $y$.
}
\label{fig:caseA2_2_tr} 
\end{figure*}

\begin{figure*}[htp] 
\centering%
\includegraphics[width=1.7\columnwidth]{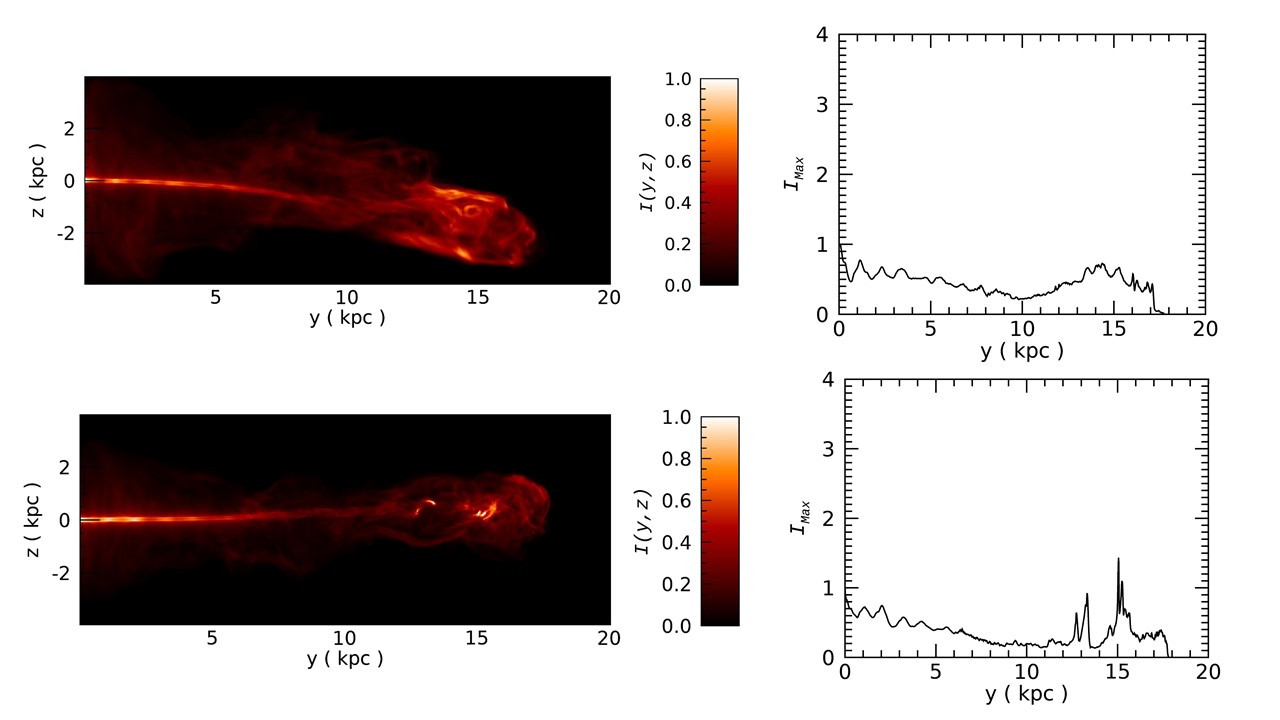}%
\caption{Top left panel: Surface brightness distribution for case A1  in the $(y,z)$ plane at the time $t=3.2 \times 10^7$ yrs. Bottom left panel: Same as    
for  case A2, but at the time $t=2.1 \times 10^7$ yrs. Right panels: Corresponding maximum brightness as a function of $y$ for case A1 (top) and for case A2 (bottom). }
\label{fig:brightA1-2} 
\end{figure*}

Overall, the velocities are weakly relativistic for  cases A1-2 and mildly relativistic for cases B and C1-2. The jet-counterjet brightness ratio corresponding to these jet velocity turns out to be compatible with the observational values \citep{odonoghue93}.
Following Paper II, in order to compare the simulation results with the radio images, we defined a synthetic surface brightness distribution ${\mathcal{I}}(x,y)$ map obtained by integrating along the line-of-sight of the 3D distribution of emissivity $\epsilon(x,y,z)$, assumed to be proportional to  $p(x, y, z)  B^2(x, y, z)$, that is to the product of the thermal pressure and the square of the magnetic field strength:

\begin{equation}
{\mathcal{I}}(x,y) = \int_{-L_z/2}^{L_z/2} \epsilon(x,y,z) \ dz \, .
\label{eq:bright}
\end{equation}
where we assumed a line-of-sight along the $z$ axis and we performed the integral over the extent of the computational box along this axis.

 The simulated jet temporal evolution and the magnetic instability impact on the attained morphologies are clearly shown by means of tracer animations accompanying the paper. We therefore recommend that the readers look at these animations for a better understanding of the complex details of the physical processes that govern the jet propagation.

\subsection{Cases A1-A2}

We began our analysis with the low-power case A1 (see Table \ref{labvalues}), with the $\eta$ parameter set to $10^{-3}$ ($=1/10$ of the value used in Paper II), the jet velocity  $V_{\mathrm{j}} =0.1$, and the jet kinetic luminosity results ${\cal L}_{\mathrm {kin}} = 7 \times 10^{42} {\mathrm {erg \ s}}^{-1}$.
In Papers I and II, we observed that during the initial phases all jets presented an FR II morphology, with a strong termination shock.
In Fig.  \ref{fig:caseA1_1-2} (top panels), we took a snapshot at $t=1.9\times 10^7$ yrs; in the top-left panel we show a cut  in the $(z,y)$  plane of the density distribution, while in the top-right panel we plotted the maximum pressure found at each $y$ along the jet, as a function of $ y$. As discussed in Papers I and II, the presence of a strong shock at the jet's head can be an indicator of an  FR~II morphology, since it can be identified with the hot spot.  From Fig. \ref{fig:caseA1_1-2}  we can notice that also in this case, in this early phase, the morphology is of FR~II type, as confirmed by both  the density distribution and by the behaviour of the maximum pressure that shows a termination shock at the jet head.

It is interesting to compare the outcomes of a different choice of the $\alpha$ parameter, which is represented by case A2. For A2 we used $\alpha = 2$, which is the same value from Paper II, and this gives a steeper density profile. Incidentally, we note that as a consequence of our choice of constant pressure in the ambient medium, the steeper decrease in density, as a function of the radial distance, corresponds to a steeper increase in temperature. However, at distances below $20$ kpc from the source, that is $< 5 \ r_ {\mathrm c}$, the ambient temperature reaches a value $ \lesssim 5$ keV, which is still consistent with values found in the Intracluster Medium (ICM) \citep{vik06}. 
The bottom panels in Fig.  \ref{fig:caseA1_1-2}  display, for case A2, the same quantities as the top panels. The snapshot was taken  at a slightly  earlier time, $t=1.3\times 10^7$ yrs, when the jet head had reached the same position as in case A1. Since the density has a steeper drop, the jet head has a larger velocity for case A2 than for case A1 and it reached  a similar position earlier. From the comparison between the top and bottom panels, we can  notice that the evolution differs only  in the details and also in this case we see a morphology typical of a small FR~II.

To follow the jet propagation one can look at the animation for case A1  ($CaseA1.mp4$) which also shows how the tracer distribution would appear as seen under different viewing angles. We can see the jet wobbling and the jet head periodically deflecting at angles that can reach 90 degrees and then it resumes its straight course. This behaviour may be attributed to current driven instabilities, as we  discuss in a more detailed way when analysing cases C1 and C2.  At later times, when the jet head reaches about 10 kpc, in case A1 we observe a large-scale bending of the jet,  which is not visible in case A2, where the jet remains straight (readers can refer to $CaseA2.mp4$ for the animation). This difference is evident by comparing Figs \ref{fig:caseA_1_tr} and \ref{fig:caseA2_2_tr}, where, in the left panels, we display a 3D rendering of the tracer distribution, at $t=3.2 \times 10^7$ yrs, for case A1, and at $t=2.1 \times 10^7$ yrs, for case A2, when the jet head has reached the end of the computational domain. The bending is clearly visible for case A1, while case A2 is characterized by a straight jet.

In the same figures in the right panel, we show 
a plot of the maximum pressure (as in Fig. \ref{fig:caseA1_1-2}).
From the plot we see that, at this epoch, the termination shock has disappeared and the jet has acquired an FR~I morphology. In case A2, however,   we can observe spikes 
at $y \approx 13 $ and $15$ kpc, which, as discussed in Paper II, could be signatures of warm spots. The jet morphology can be better appreciated by looking at Fig. \ref{fig:brightA1-2}, where, on the left, we show a synthetic  surface brightness distribution for both cases (A1, top panel and A2, bottom panel) and, on the right, a profile of the maximum brightness as a function of $y$. Hereinafter, these images were obtained by computing the quantity $\mathcal{I}(x,y)$ defined by Eq. \ref{eq:bright}, after normalizing the brightness amplitude by dividing by the value assumed at the point of coordinates $(0,0)$. In both cases the jets are well visible and terminate in fainter lobes with no clear hot spots. In the lobes however, regions of higher emission are present. The higher emission ridge in case A1 outlines the region where the jet bending is more pronounced (redaders can compare this with Fig. \ref{fig:caseA_1_tr}), while the bright spots in case A2 correspond to the peaks in the right panel of Fig.  \ref{fig:caseA2_2_tr}. The brightness profile for case A1, on the right, shows very small variations, with a slight decrease along the jet and a slight increase in correspondence of the lobe. In case A2, instead, we observed  two spots of increased brightness in the middle of the lobes which, as already observed, can be reminiscent of warm spots marking the transition to faint lobes, similarly to what we observed in Paper II.

\subsection{Case B}

\begin{figure*}[htp] 
\centering%
\includegraphics[width=1.6\columnwidth]{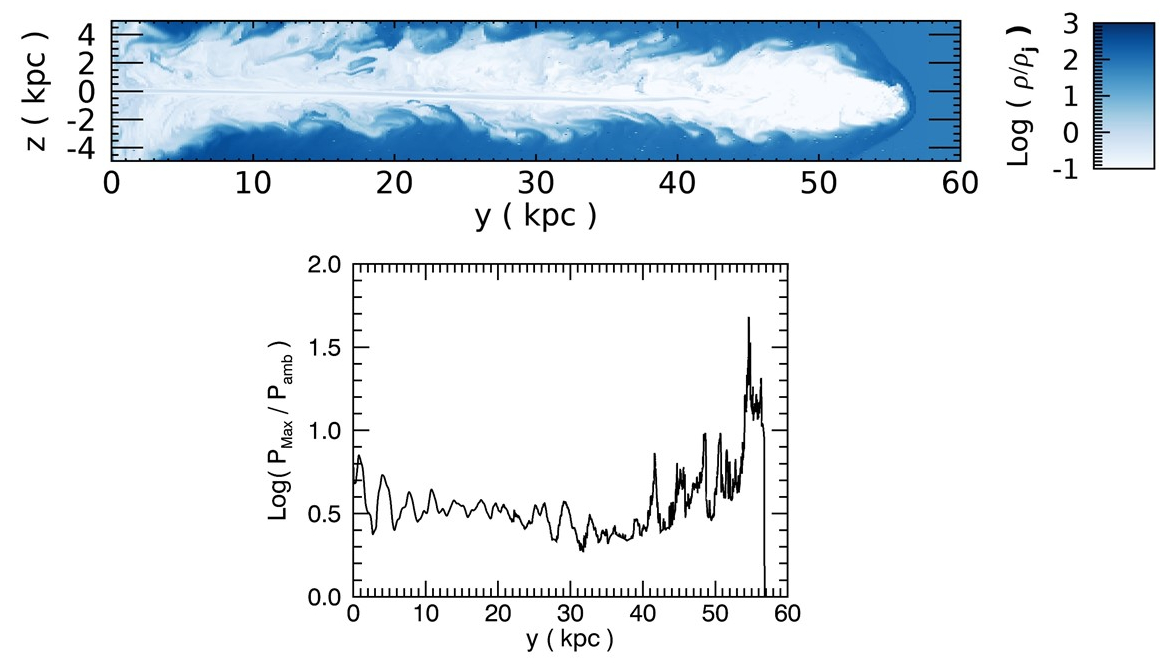}%
\caption{\footnotesize  Top: Cut of the logarithmic density distribution  in the $(y,z)$ plane for the high luminosity case B at $t\approx 1.8 \times 10^7$ yrs.  Bottom: Logarithm of the maximum gas pressure as a function of $y$.
}
\label{fig:caseB} 
\end{figure*}

\begin{figure*}[htp] 
\centering%
\includegraphics[width=1.6\columnwidth]{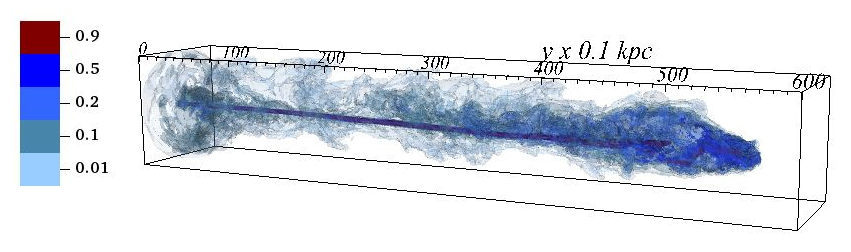}%
\caption{\footnotesize 3D isocontours of the tracer distribution for case B at $t\approx 1.8 \times 10^7$ yrs. The size of the computational box is $10 \times 60 \times 10$ kpc. An FR~II morphology is displayed.
}
\label{fig:tracer_3D_B} 
\end{figure*}

\begin{figure*}[htp] 
\centering%
\includegraphics[width=1.6\columnwidth]{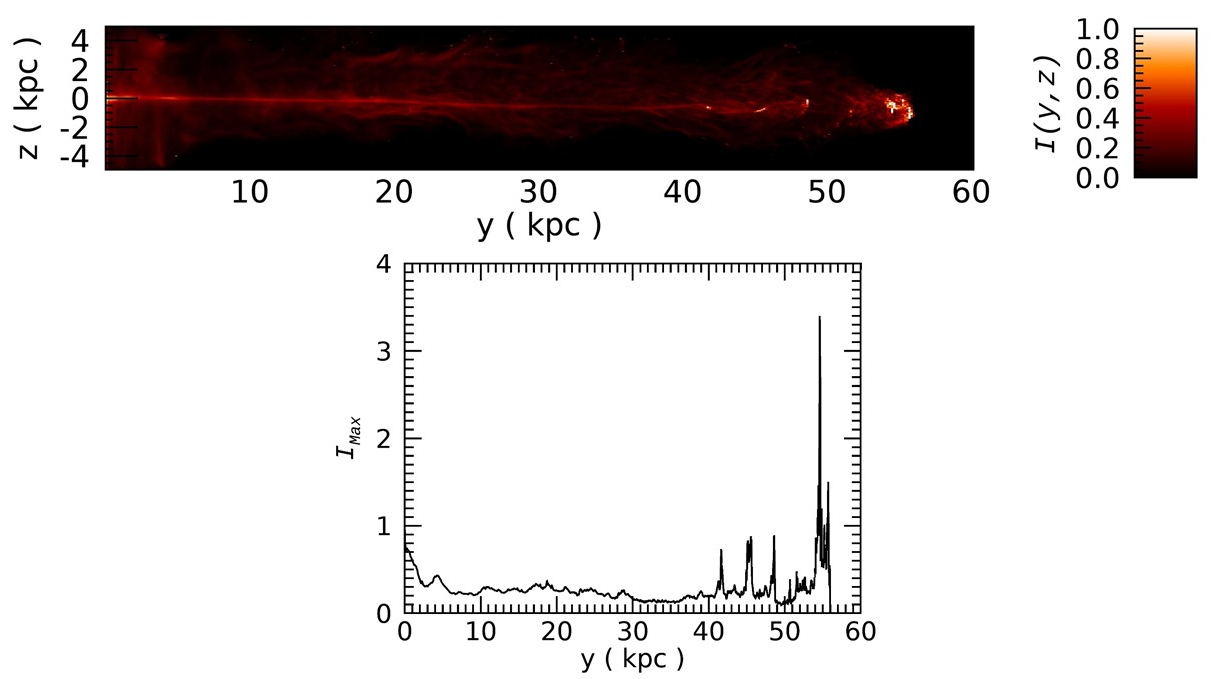}%
\caption{\footnotesize  Top: Brightness distribution in the $(y,z)$ plane in  case B at $t\approx1.8 \times  10^7$ yrs.  Bottom:  Corresponding maximum as a function of $y$.
}
\label{fig:BB_caseB} 
\end{figure*}

We now turn our attention to Case B, which has a power about 30 times higher than the previous Case A. In Fig. \ref{fig:caseB} we show,  in the top panel,  a longitudinal cut of the density distribution in the plane $(y,z)$  and, in the bottom the maximum pressure. We notice that the spatial extension of the plots is three times larger than in the previous cases A1 and A2 and the jet head  reached the end of the grid at about $60$ kpc. The snapshot was taken at $t \approx 1.8 \times 10^7 \; \hbox{yrs}$, which is approximately equal to the time of the snapshots in  Fig. \ref{fig:caseA1_1-2} when the jet  reached a length of about $15$ kpc. The jet head in cases B, therefore, propagates at a mean velocity which is about four times larger than in cases A1 and A2. Additionally the jet propagates straight with little wobbling and no bending up to the final stages of the evolution (see animation in $CaseB.mp4$) when non-linear transverse modes begin to show their effects. This is confirmed by Fig. \ref{fig:tracer_3D_B}, where we show  3D isocontours of the tracer distribution at the same time.  Therefore, in this case, the MHD instabilities have weaker effects on the jet propagation in the short and medium period.  In the  plot of the maximum pressure (bottom panel), we can observe  the presence of a strong termination shock,  located at about $55$ kpc, preceded by a series of peaks of a much lower strength.  In Fig. \ref{fig:BB_caseB} in the plane  $(y,z)$, we show the surface brightness distributions. Consistently to what was discussed before, the surface brightness distribution shows the presence of a hot spot located at about $55$ kpc corresponding to the so called splash point, where the jet terminates into the IGM.  This lead us to conclude that jets with this set of parameters would exhibit an FR~II morphology.

\subsection{Case C1-C2}

In cases C1 and C2, we decreased $\eta$ by about a factor of 3 with respect to case B and, accordingly, the jet power, which turns out to be intermediate between cases A and B. As for the A cases, we considered two different density stratification parameters: we used $\alpha = 1$ for case C1 and $\alpha = 2$, and thus a steeper gradient, in  case C2.  We also note that the extension of the grid along the jet direction is $60$ kpc as for case B, but it is much longer than for the A cases. Since we assumed a constant pressure, the decrease in density is  compensated for by an increase in temperature  and, as a consequence, in Case C2 we reached very high temperatures that are not compatible with observational data. Notwithstanding this unrealistic character, case C2 can provide hints about the effects of a steeper density stratification.

\begin{figure*}[htp] 
\centering%
\includegraphics[width=1.8\columnwidth]{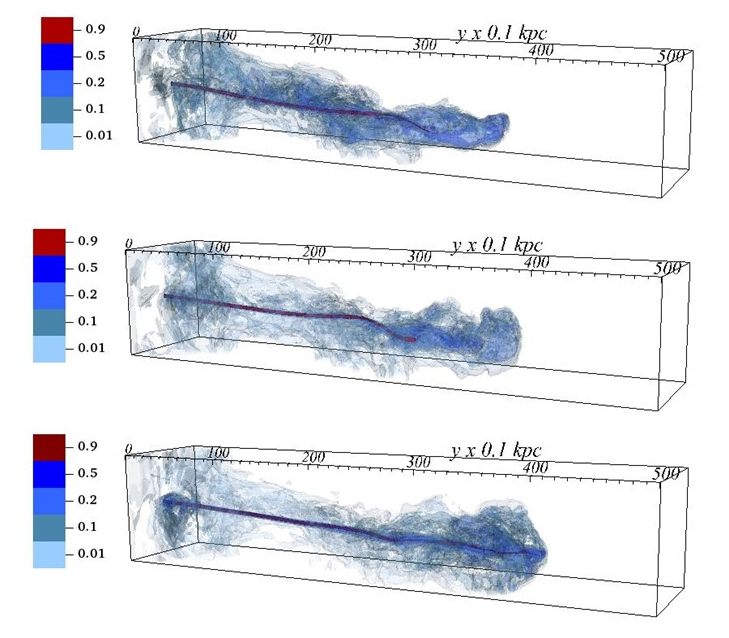}%
\caption{3D isocontours of the tracer distribution for case C1 at three different times, in the top panel $t \sim 3.7 \times 10^7 $ yrs, in the middle panel $t  \sim  4.3  \times 10^7 $  yrs and in the bottom panel $t \sim  5.1 \times 10^7$ yrs.
}
\label{fig:caseC1} 
\end{figure*}

\begin{figure*}[htp] 
\centering%
\includegraphics[width=1.8\columnwidth]{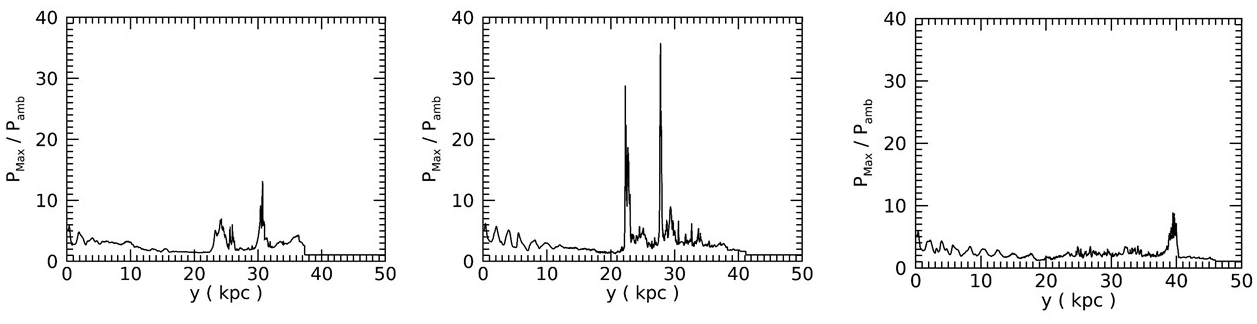}%
\caption{Plot of the maximum pressure as a function of $y$ for case C1 at three different times, which are the same as in the previous figure, namely in the left panel $t \sim 3.7 \times 10^7 $ yrs, in the middle panel $t  \sim  4.3  \times 10^7 $ yrs and in the right panel $t \sim  5.1 \times 10^7$ yrs.
}
\label{fig:pmax_C1} 
\end{figure*}

\begin{figure*}[htp] 
\centering%
\includegraphics[width=1.8\columnwidth]{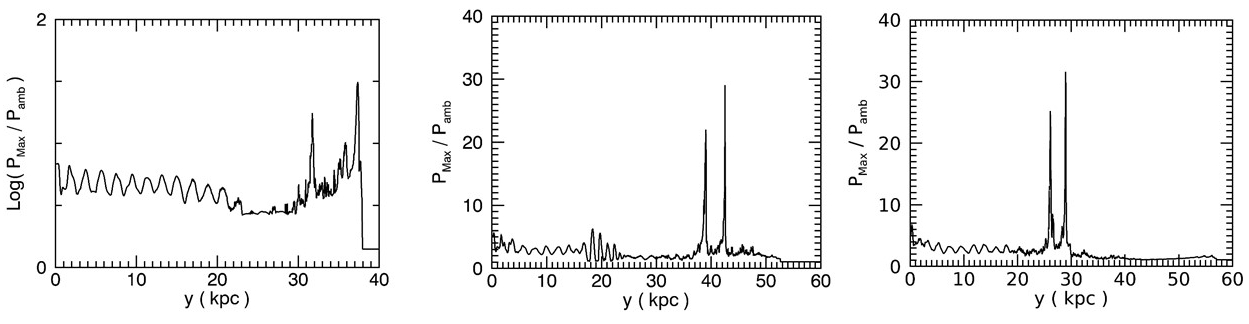}%
\caption{Plot of the maximum pressure as a function of $y$ for case C2 at three different times, in the left panel $t \sim 1.4 \times 10^7 $ yrs (in logarithmic scale), in the middle panel $t  \sim  2  \times 10^7 $ yrs and in the right panel $t \sim  2.2 \times 10^7$ yrs.
}
\label{fig:pmax_C2} 
\end{figure*}

\begin{figure*}[htp] 
\centering%
\includegraphics[width=1.8\columnwidth]{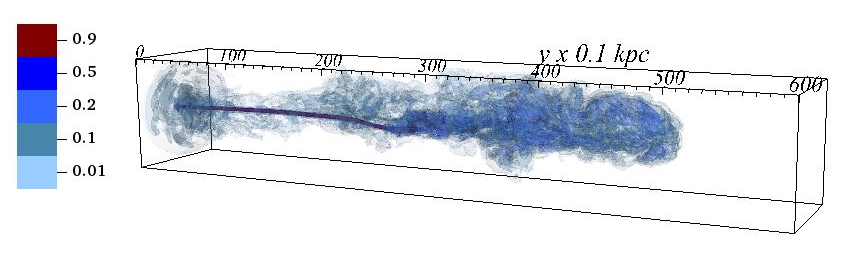}%
\caption{3D isocontours of the tracer distribution for case C2 at $t=2.2 \times 10^7$ yrs. The size of the computational box is $10 \times 60 \times 10$ kpc.
}
\label{fig:CaseC2} 
\end{figure*}

\begin{figure*}[htp] 
\begin{center} 
\includegraphics[width=1.8\columnwidth]{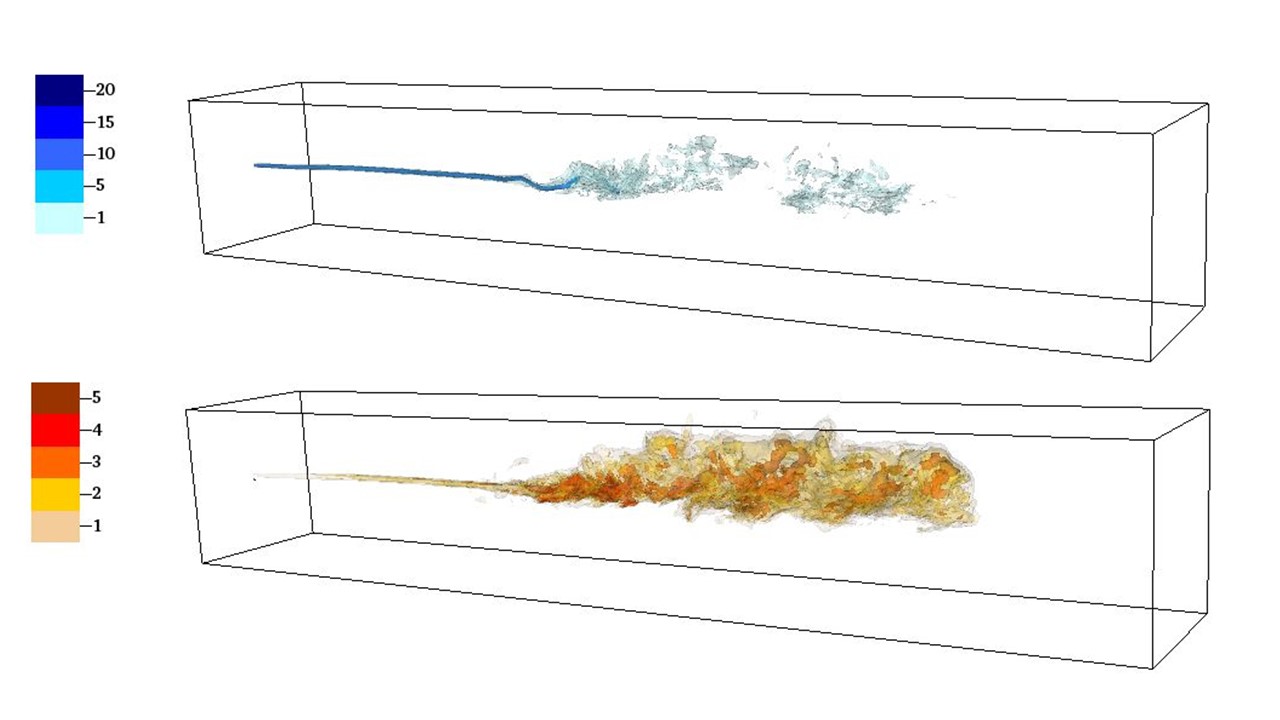} 
\end{center} 
\caption{3D isocontours of the longitudinal momentum density (top) and of the specific thermal energy (bottom) distributions for case C2 (in arbitrary units), at $t=2.2 \times 10^7$ yrs. }
\label{fig:mom-temp-3D} 
\end{figure*}

\begin{figure*}[htp] 
\centering%
\includegraphics[width=1.8\columnwidth]{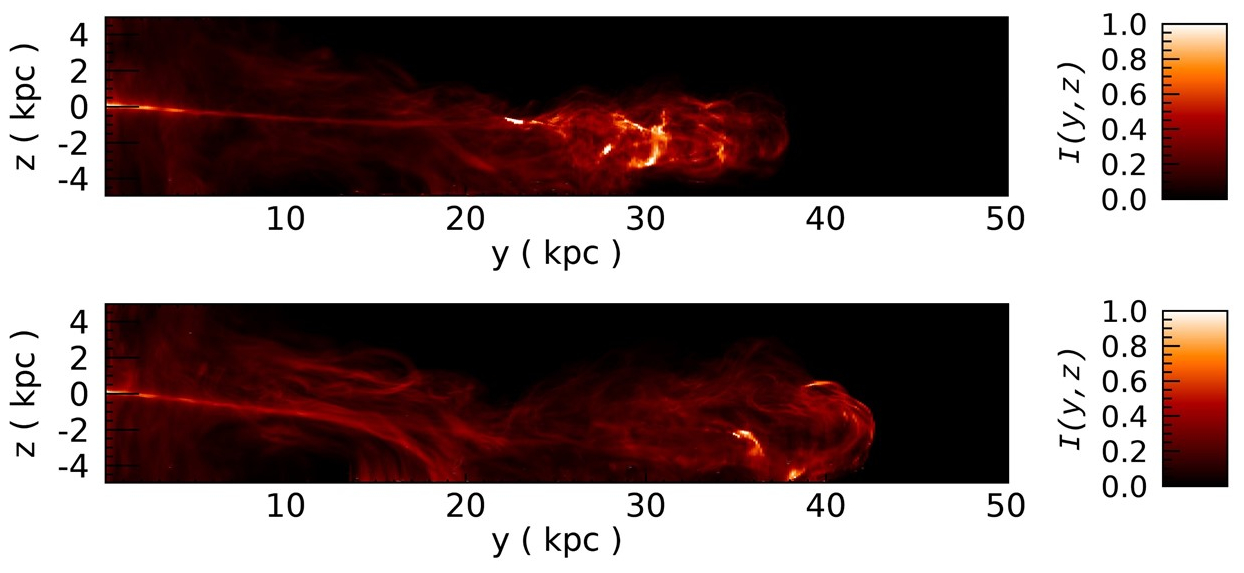}   \\

\caption{Surface brightness distribution for case C1  at  $t=4.3 \times 10^7$ yrs (top panel) and $t = 5.1 \times 10^7$ yrs (bottom panel). 
}
\label{fig:brightC} 
\end{figure*}

We can follow the temporal jet evolution by looking at the animation for  case C1 ($CaseC1.mp4$) which, as in the case A1, shows the appearance of the tracer of the bottom panel of Fig. \ref{fig:caseC1} under different viewing angles. As in Cases A1-2, we observed jet wobbling and periodic deflections at angles that can reach 90 degrees.  When this happens, the jet may break and, after the point of breaking,  we observed a slower,  turbulent,  and much more diffused flow, with a strong mixing with the ambient medium. However, after some time, the jet is able to regain its straight course.  The interval of time during which the jet appears to be broken increases as the jet length increases.  We can look in more detail at this breaking and restarting phase for case C1 in Fig. \ref{fig:caseC1}  where we show  a visualization of the 3D tracer distribution at three different times, and in Fig.  \ref{fig:pmax_C1} where we show the plot of the maximum pressure for the same three times.  The images of the tracer distribution in the top and middle panels of Fig. \ref{fig:caseC1}, which  are  at $t  \sim 3.7 \times 10^7 $ yrs  (top panel) and $t  \sim  4.3  \times 10^7 $ yrs (middle panel), respectively, show the collimated jet that reaches  a distance of about 30 kpc in both panels. At distances larger than 30 kpc, we can observe  a diffused region, which is much more mixed with the ambient medium, as indicated by the light blue colour. The length of this region increases from the top to the middle panel; in fact, it reaches about 38 kpc in the top panel and about 41 kpc in the middle panel. The  bottom panel at $t \sim  5.1 \times 10^7$ yrs shows the restarting of the jet which appears to be well-collimated  up to the distance of 41 kpc, which was reached only by the mixed and diffused portion earlier. We  thus have an estimate of the temporal length during which the jet was broken off: about $10^7$  yrs.  The same phases can be observed in the maximum pressure plot; the point of the collimated jet breaking is marked by the peaks between 20 and 30 kpc, visible in both the left and middle panels. In the right panel those peaks disappear and a new peak, although much weaker, appears at about  the distance reached by the collimated portion of the jet (40 kpc).

A similar behaviour can also be observed in Case C2: in Fig. \ref{fig:pmax_C2} we show the maximum pressure behaviour along the jet at three different epochs. The first plot is at $t \sim  1.4 \times10^7$  yrs  (it is important to note that the left panel shows the logarithm of the pressure): we can observe a strong shock at the jet  head and we note the formation of a strong secondary shock at about 32 kpc, but still we have an FR II-like appearance. The middle plot is at  $t \sim  2 \times 10^7$  yrs,  the head shock has disappeared, while two weaker shocks  are visible at $y \sim 38 -42 $kpc and they represent the transition from the  collimated portion of the jet to the more diffused and turbulent portion. At  $t \sim 2.2 \times 10^7$  yrs (right panel) the two shocks  positioned at $y \sim 26 -29$ kpc, while the diffused portion of the jet reaches  $y  \sim 55$ kpc.  The appearance of the jet at the last time is visible in Fig. \ref{fig:CaseC2}, where we show the 3D tracer distribution; in the figure we can observe the collimated portion of the jet that reaches about half of the total jet length which is followed by the more diffused and  uncollimated region. The breaking of the collimated jet can be better appreciated in Fig. \ref{fig:mom-temp-3D} as well as in the animation ($CaseC2.mp4$). In the top panel  we display the 3D distributions of the longitudinal component of the (mechanical) momentum density of the jet which, at  $y \sim 26 - 29$  kpc,  shows an abrupt transition, where the jet dissipates a relevant fraction of its mechanical momentum and kinetic power into thermal energy. The effect of this dissipation is visible in the bottom panel, where we display  the specific thermal energy  that  shows a substantial  increase at the same position.

The difference between the two cases C1 and C2 is that, after the breaking, the diffused portion of the jet reaches, in a similar time, a much larger distance in Case C2 than in Case C1. In fact, the length of this portion is about 12 kpc in case C1 and about 25 kpc in case C2. This effect may be explained by the steeper decrease of the density in the ambient medium in Case C2 with respect to case C1. In case C2 the restarting of the collimated jet had not occurred yet at the final time of the simulation, but, based on the results of Case C1, we can expect that at later times it will occur. In general, all the morphologies that we have described so far have to be considered as transient, with timescales of variation that can reach $\sim 10^7$yrs. 

In Fig. \ref{fig:brightC} we show, for case C1, the `proxy' surface brightness distribution at two different times,   $t=4.3 \times 10^7$ yrs  (top panel) and $t=5.1 \times 10^7$ yrs (bottom panel), corresponding  to the top and bottom panels of Fig. \ref{fig:caseC1}, respectively. Comparing the two panels, we can observe  a substantial change in the morphology:  in the top panel we have a `warm spot' around $y \sim 23$ kpc, almost at the end of the collimated portion of the jet, followed by a diffused lobe, in the bottom panel the morphology is of FR~II type with the hot spot at the jet termination.  This result again  underlines the very dynamic behaviour of this jet, with transient changes of  morphologies that can range from FR~I to FR~II types.

\subsection{Case C-RHD}

\begin{figure*}[htp] 
\centering%
\includegraphics[width=1.8\columnwidth]{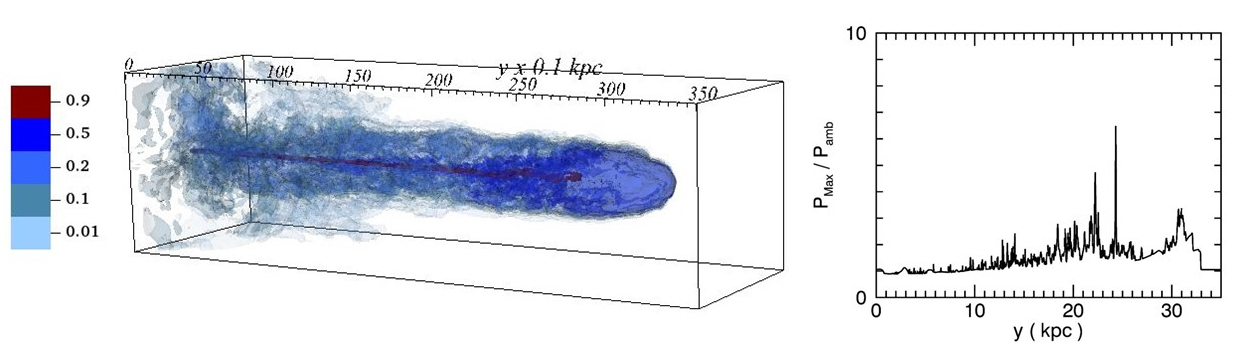}%

\caption{Left: 3D isocontours of the tracer distribution for case C-RHD, at $t=3.6 \times 10^7$ yrs; the size of the computational box is again $10 \times 35 \times 10$ kpc. Right: Maximum gas pressure as a function of $y$ at the same time.
}
\label{fig:caseRHD2} 
\end{figure*}

\begin{figure*}[htp] 
\centering%
\includegraphics[width=1.8\columnwidth]{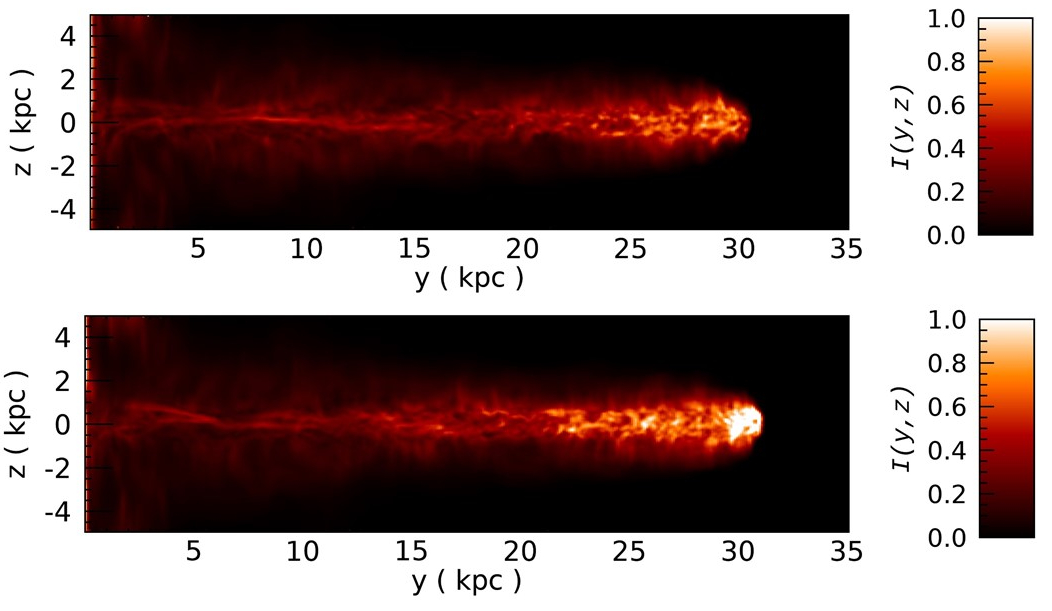}%

\caption{Brightness distribution in the $(y,z)$ plane for  case C-RHD at two different times: $t\approx3.4 \times  10^7$ yrs (top panel) and $t\approx3.6 \times  10^7$ yrs (bottom panel).
}
\label{fig:bright_C-RHD} 
\end{figure*}

To better understand the role of the magnetic field in the jet propagation, we carried out a simulation adopting the same parameters of Case C1, but in the RHD limit, that is taking a value of $\beta = 10^6$. The animation shows that, differently from Case C1, the jet propagates straight, without any wobbling or bending. As in the previous cases, in the early phases the jet is characterized by a strong termination shock, therefore maintaining an FR II-like morphology. As the jet advances, however, it is slowed down by a series of shocks well ahead of its termination. This is clearly depicted by Fig. \ref{fig:caseRHD2} which, at $t=3.6 \times 10^7$ yrs, displays   a 3D  rendering of the tracer distribution  in the left panel and, in  the right panel, a plot of the maximum pressure. In this phase,  the strong termination shock disappears and the jet deceleration occurs through a series of weaker shocks, still located in the termination part of the jet. The straight propagation of the jet and the absence of bendings can be connected to the very low magnetization of this case, in which the magnetic field can be considered essentially absent. In fact, the magnetization strength is the only difference between this case and case C1, in which the jet wobbling can be attributed to the effects of current driven instabilities. In Fig. \ref{fig:bright_C-RHD} we show the synthetic brightness distribution at two different epochs: at $t\approx3.4 \times  10^7$ yrs (top panel) and $t\approx3.6 \times  10^7$ yrs (bottom panel), the final time of the simulation. We see that at a temporal distance of about two million years, a source would be classified as FR~I at the earlier time and, at the later time, being  
image-dominated by the strong emission in the terminal part of the jet, the same source would look like an FR~II. We may however speculate that the jet will progressively decelerate and may at later times, acquire an FRI morphological character.

\begin{figure*}[htp] 
\centering%
\includegraphics[width=1.8\columnwidth]{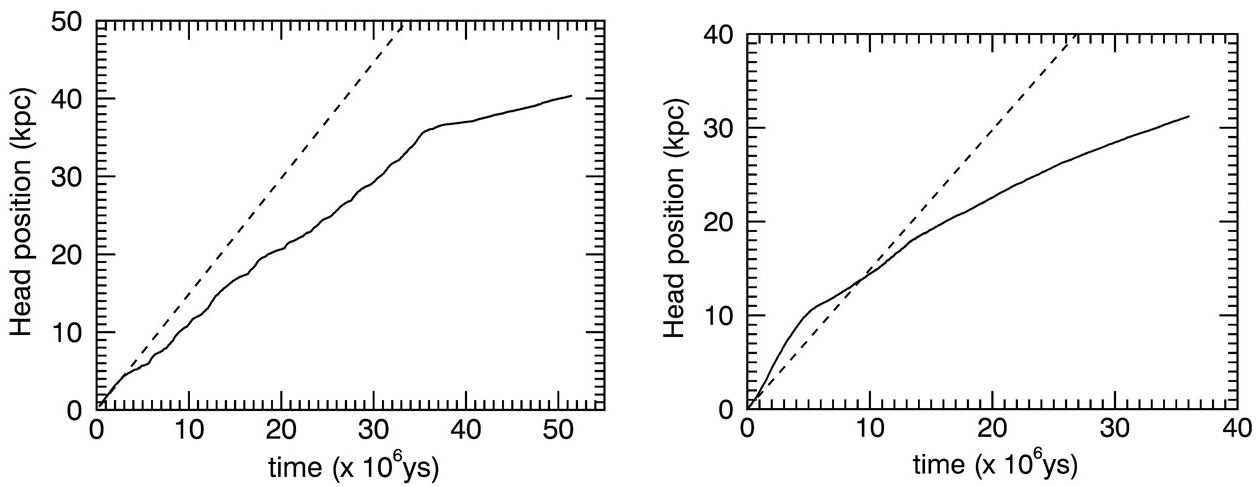}%

\caption{ Location of the jet's head as a function of time for case C1 (left panel) and C-RHD (right panel). 
}
\label{fig:yhC} 
\end{figure*}

It is also interesting to contrast the different behaviours of the jet head propagation in the two cases C1 and C-RHD. In Fig. \ref{fig:yhC} we plotted the jet's head location as a function of time, the plot in the left panel refers to Case C1 and that in the right panel refers to case C-RHD. The solid line represents the actual head's position while the dashed line corresponds to the 1D approximation based on the momentum conservation \citep{marti97}. In the case C1, the solid line is always below the dashed one and the slope of the curve decreases abruptly at about $3.5 \times 10^7$ yrs, when, as we discussed above, we observed a jet breaking with a transition to an uncollimated flow.  Conversely, in the C-RHD case, up to about $10^7$ yrs the actual head position is above the 1D limit as a result of the `beam pumping' (see \cite{ksl88,Massaglia96, belan13}) which sets in when the jet maintains its axial symmetry. When the 3D, non-axial effects become dominant the head velocity decreases and the slope of its position curve declines smoothly. This smooth decrease in the jet head velocity can also be clearly appreciated by looking at the animation.

%%%%%%%%%%%%%%%%%%%%%%%%%%%%%%%%%%%%%%%%%%%%%%%%%%%
\section{Discussion and conclusions}
%
%
%%%%%%%%%%%%%%%%%%%%%%%%%%%%%%%%%%%%%%%%%%%%%%%%%%%

We presented 3D relativistic magneto-hydrodynamic simulations of jets propagating into a stratified medium. The cases that we considered differ by their velocity and density contrast with the ambient  and, therefore, by their powers.  We considered powers that differ by about one order of magnitude and that are at the border between FR I and FR II and we also investigated the effects of different density stratifications.  During their evolution, the jets take a multitude of variegated morphologies, which depend on  physical parameters, but they may also have a transient character. In particular,  we have shown that, besides the density ratio and the Lorentz factor, the magnetic field  plays an important role in shaping these morphologies.

In all cases the jets produced  an FR~II-like morphology during the early stages of the propagation, up to distances $10-20$ kpc from the injection boundary, which is in agreement with Papers I and II. Subsequently, the following evolution diverges, depending on the parameters.

In more detail, the first two simulations, cases A1 and A2, have a similar (low) kinetic power of the MHD cases of Paper II, but with a lower density ratio and a higher velocity. The evolution is very  similar to that presented in Paper II; the jet is disrupted by magnetic instabilities at a distance of about $12-14$ kpc, the flow is strongly decelerated, and it becomes highly turbulent. The differences in the density stratifications lead to some changes in the morphological details, but the general behaviour given by the surface brightness distributions points towards an FR~I regime.   

The high power Case B maintains an FR~II morphology during  its entire evolution, with a hot spot at the jet's head up to the last frame at $55$ kpc. This was pointed out by the behaviour of the maximum pressure and emissivity plots and by the surface brightness distribution that clearly shows the presence of the terminal hot spot.

Cases C1 and C2 have an intermediate power between  cases A and B and, as in cases A, the magnetic field plays a fundamental role in the jet dynamics, leading, through the development of current driven instabilities, to jet wobbling, bending and breaking. The role of the magnetic field is demonstrated by a comparison between cases C1-2 and C-RHD, in which the magnetic field is negligible. In fact,  in this last case, in  contrast to the other two cases,  the jet propagation direction remains straight. As already mentioned, in cases C1-2,  MHD instabilities may eventually lead to  jet breaking and  a transition from a collimated jet to a turbulent much slower flow, with velocities that, at most, reach $\sim 1,000 \; \mathrm{km \ s}^{-1}$. The transition is marked by a warm spot, followed by a diffuse lobe of emission. This configuration, however is transient and after about $10^7$ yrs from the jet breaking the jet restarts and regains its straight course, while the morphology takes the aspect of a FR~II source.

From a more general point of view, the present results, together with the findings in Papers I and II show that the radio source morphologies do not depend  on the jet luminosity only: density ratios and magnetic fields play an important role as well. For example, from jets of a similar power one can obtain either an FR~I morphology  or a WAT morphology depending on the magnetic field strength and, as shown in Paper II, the possible  relative motion of the source in the IGM. In addition, for intermediate jet powers, different kinds of morphologies are obtained during the evolution at different times.  The role of the magnetic field in inducing jet deflections  has  also been demonstrated by \cite{Mig2010} who carried out numerical simulations of highly relativistic and strongly magnetized jets propagating in a uniform ambient medium and by \cite{mukh20} who studied the evolution of magnetized jets with Lorentz factors 3-10 propagating in a stratified medium. Albeit differing very much in the parameters and in the properties of the ambient medium, with respect to the simulations presented in this paper, both of these analyses showed morphologies characterized by wiggles and wobbles of the jets, more or less intense, as in the present paper.

The morphologies derived from the numerical experiment have strong similarities with the observed brightness distribution of FR~II radio galaxies that often show a wealth of structures besides the hot spots and the symmetric radio lobes. In order to associate real radio sources with the results of the simulations we explored the morphologies of those included in the Third Cambridge Catalogue \citep{spinrad85}. We limited our analysis to the 3C sources with redshift z$\lesssim$0.5 in order to maintain a good spatial resolution, searching for images in the NRAO VLA Archive Survey Images Page, available at \footnote{\sl http://archive.nrao.edu/nvas/}. 

\begin{figure*}[htp] 
\centering%
\includegraphics[width=1.6\columnwidth]{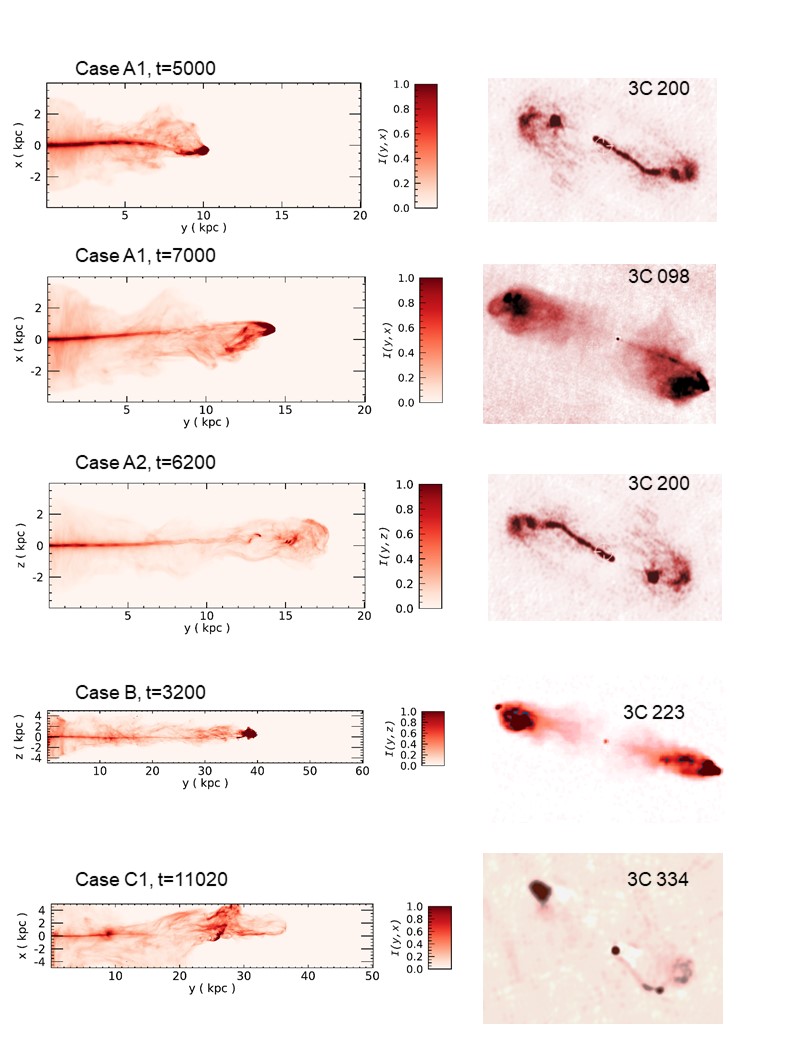}%
\caption{\footnotesize  Simulated brightness distribution for different cases at different epochs (left column), compared with radio maps for sources of the 3C Catalogue (right column).
}
\label{fig:cartoon} 
\end{figure*}

In Fig. \ref{fig:cartoon} we show a comparison of a sub-set of five simulated brightness distributions with those observed in the radio band (in some cases the radio maps have been rotated by a multiple of 90 degrees for an easier comparison). All of these images show distortions close to the jet end, closely reproduced in the radio maps. For example, in 3C~200 (top panel), the jet bends and it is followed by multiple bright regions. In 3C~098 (second panel), the plasma flowing back towards the nucleus is highly asymmetric, following the bend at the jet's end. We note that these two objects are related to the same simulation (case A1) just at different times with  strongly different morphologies, result that stresses the importance of time evolution. In the other sources, we noticed warm spots located along the jet path, while in the source (3C~334) the jet produces a `hook-like' structure, also observed in the simulations. 

In synthesis: high power jets lead to FR~II radio sources and low power jets lead to FR~Is, while jets of intermediate power may appear to observations as FR~Is of FR~IIs depending on density ratios, and magnetic field intensity. We also stress that, because many features are highly time-dependent, the observation epoch also plays an important role.
These results provide a perspective on future work, to analyse how jets of the same kinetic power evolve into different morphologies depending upon their values for the magnetization and density ratio.

\begin{acknowledgements}
 We acknowledge the computing centre  of Cineca and INAF, under the coordination of the \it{Accordo Quadro MoU per lo svolgimento di attivit\'a congiunta di ricerca Nuove frontiere in Astrofisica: HPC e Data Exploration di nuova generazione}, for the availability of computing resources and support. We acknowledge also support from {\it{PRIN MIUR 2015}} (grant number 2015L5EE2Y). Calculations have been also carried out at the  Competence Centre for Scientific Computing (C3S) of the University of Torino. We are also indebted to an anonymous reviewer whose competent and careful objections lead us to radically revise and improve the paper.
 \end{acknowledgements}

\bibliographystyle{aa}
\bibliography{main.bib}

\label{lastpage}
\end{document}